\documentclass[12pt]{article}
\usepackage{multirow}
\usepackage{verbatim}

\usepackage{authblk} 
\usepackage{float} 
\usepackage{amsmath,amsthm,amssymb,bbm,mathrsfs,mathtools,xfrac} 

\usepackage{adjustbox}
\usepackage[sort]{natbib}   
\usepackage{placeins} 
\usepackage[pagebackref=true,bookmarks]{hyperref}
\hypersetup{
	unicode=false,
	pdftoolbar=true,
	pdfmenubar=true,
	pdffitwindow=false,     
	pdfstartview={FitH},    
	pdfkeywords={}, 
	pdfnewwindow=true,      
	colorlinks=true,       
	linkcolor=red,          
	citecolor=blue,        
	filecolor=black,      
	urlcolor=cyan           
}
\usepackage[utf8]{inputenc} 
\usepackage[T1]{fontenc} 
\usepackage{ctable} 
\usepackage{pifont}

\usepackage{array}
\newcolumntype{L}{>{\centering\arraybackslash}m{3cm}} 
\usepackage{color, colortbl, xcolor, comment}
\usepackage{subfig}
\usepackage{threeparttable}
\usepackage{tcolorbox} 
\usepackage{algorithm}
\usepackage{chngcntr} 
\usepackage[noend]{algpseudocode}
\algrenewcommand\textproc{}
\algdef{SE}[SUBALG]{Indent}{EndIndent}{}{\algorithmicend\ }%
\algtext*{Indent}
\algtext*{EndIndent}
\usepackage{pdflscape}

\newtheorem{theorem}{Theorem}

\newtheorem*{remark}{Remark}
\newtheorem{corollary}{Corollary}

\newcommand{\bx}{\textbf{\emph{x}}}
\newcommand{\by}{\textbf{\emph{y}}}
\newcommand{\bX}{\textbf{\emph{X}}}
\newcommand{\bW}{\textbf{\emph{W}}}
\newcommand{\bY}{\textbf{\emph{Y}}}

\newcommand{\bV}{\textbf{\emph{V}}}

\newcommand{\bpsi}{\boldsymbol{\psi}}

\newcommand{\btau}{\boldsymbol{\tau}}

\definecolor{Gray}{gray}{0.9}

\newcommand{\bbeta}{\boldsymbol{\beta}}
\newcommand{\bR}{\boldsymbol{R}}
\newcommand{\bA}{\boldsymbol{A}}

\newcommand{\btheta}{\boldsymbol{\theta}}

\DeclareMathOperator*{\argmin}{arg\,min}

\DeclareMathOperator{\diag}{diag} 

\DeclarePairedDelimiter\abs{\lvert}{\rvert}%
\DeclarePairedDelimiter\norm{\lVert}{\rVert}%

\newcommand{\ba}{\boldsymbol{a}}
\newcommand{\bh}{\boldsymbol{h}}

\makeatletter
\let\oldabs\abs
\def\abs{\@ifstar{\oldabs}{\oldabs*}}
\let\oldnorm\norm
\def\norm{\@ifstar{\oldnorm}{\oldnorm*}}
\makeatother

\usepackage{lastpage}
\usepackage{fancyhdr}
\usepackage[parfill]{parskip} 
\usepackage[margin=1in]{geometry}
\usepackage{setspace}
\title{Variable Selection in Regression-based Estimation of Dynamic Treatment Regimes}

\author{Zeyu Bian$^{1}$\footnote{zeyu.bian@mail.mcgill.ca}, 
Erica EM Moodie$^1$, Susan M Shortreed$^{2, 3}$ and Sahir Bhatnagar$^{1, 4}$ \\
$^1$Department of Epidemiology, Biostatistics and Occupational Health, McGill University\\ $^2$Kaiser Permanente Washington Health Research Institute \\ $^3$Department of Biostatistics, University of Washington \\ $^4$Department of Diagnostic Radiology, McGill University}

\begin{document}

\date{}
\maketitle


Dynamic treatment regimes (DTRs) consist of a sequence of decision rules, one per stage of intervention, that aim to recommend effective treatments for individual patients according to patient information history. DTRs can be estimated from models which include interactions between treatment and a (typically small) number of covariates which are often chosen a priori. However, with increasingly large and complex data being collected, it can be difficult to know which prognostic factors might be relevant in the treatment rule. Therefore, a more data-driven approach to select these covariates might improve the estimated decision rules and simplify models to make them easier to interpret. We propose a variable selection method for DTR estimation using penalized dynamic weighted least squares. Our method has the strong heredity property, that is, an interaction term can be included in the model only if the corresponding main terms have also been selected. We show our method has both the double robustness property and the oracle property theoretically; and the newly proposed method compares favorably with other variable selection approaches in numerical studies. We further illustrate the proposed method on data from the Sequenced Treatment Alternatives to Relieve Depression study.


\maketitle

\section{Introduction}

Dynamic treatment regimes (DTRs) \citep{DTRs}, or adaptive treatment strategies, consist of a sequence of decision rules that aim to improve individual patients’ health outcomes by tailoring medical treatment to each patient’s information. Statistical methods can be used to identify optimal DTRs, constructing treatment rules tailored over time to individual’s information that can optimize the expected patient outcome.

DTRs can be estimated from models that include interactions between treatment and covariates, which are often chosen a priori. However, with many covariates and a complex disease process, for which competing treatment choices have heterogeneous effects, it is difficult to know which prognostic factors might be considered relevant in the treatment rule. A more data-driven approach of selecting these covariates might improve the estimated decision rules and simplify models to improve tractability. We are motivated by the Sequenced Treatment Alternatives to Relieve Depression (STAR*D) study \citep{stard}, a randomized multistage trial that aimed to determine optimal treatments  for patients with major depressive disorder. With many of covariates such as demographic and clinical characteristics collected throughout the study, it is challenging to select covariates useful for tailoring treatment from among so many based on expert knowledge only. Thus, variable selection with the objective of optimizing individualized treatment decisions becomes important.

Much of the DTR literature focuses on estimation; variable selection with the objective of optimizing treatment decisions has been considered only occasionally. \citet{jizhu} proposed a ranking method for variable selection in DTRs. Based on this approach, \citet{sas} developed the sequential advantage selection approach, which considers variables already in the model when deciding whether to include a new variable by the additional improvement provided by this variable. \citet{LASSOq2} adopted adaptive LASSO \citep{adaptive} in the context of A-learning \citep{Murphy}, \citet{al} proposed a method which used the Dantzig selector directly to penalize the estimating equations of A-learning and has the double robust property, that is, the estimators are consistent if either one of two nuisance models is correct. The topic of variable selection in a general (not DTR) context has seen many innovations \citep[e.g.,][]{LASSO, SCAD}. \citet{jizhu} noted that most variable selection approaches focus on predictive performance, and thus may not perform well in DTRs as these techniques may underestimate the importance of variables that have small predictive ability but that play a significant role in decision making. 

In this article, we follow the DTR estimation approach of dynamic ordinary least squares regression (dWOLS) introduced by \citet{wallace2015doubly}, an approach which requires only some minor pre-computation and the implementation of standard weighted regression. While having similarities to both Q-learning \citep{ql} and G-estimation \citep{gest}, it provides simplicity and intuitiveness similar to the former and benefits from the double robustness of the latter although it is suitable only for linear decision rules. By adding two penalty terms in the dWOLS model, we perform estimation and variable selection for DTRs simultaneously. The rest of this article is organized as follows. In Section 2, we introduce the proposed penalized dWOLS (pdWOLS) approach, followed by algorithmic details and theoretical properties. Three simulation studies are given in Section 3. Finally, we apply our method to the STAR*D trial data in Section 4.

\section{Methodology}
\subsection{Introductory Concepts and Notation}
We make assumptions to proceed with estimation of DTRs: (1) Stable unit treatment value assumption (SUTVA) \citep{rubin1980}: a patient’s potential outcome is not affected by other patients’ treatment assignments. (2) Ignorability: ignorability or no unmeasured confounding \citep{ignore} specifies that for any possible treatment regimes, the stage $k$ treatment is independent of future potential covariates or outcome conditional on current patient history. (3) No interference, no measurement error, and all the individuals have complete follow-up. 

We adopt the setup of \citet{wallace2015doubly}. For a $K$-stages DTR, the following notation is used, with lowercase being used for observed variables and uppercase for their random counterparts:  $y$ denotes patient outcome (continuous) which is measured at one point in time. The goal of DTRs is to make treatment decisions that can optimize (typically, maximize) the outcome. The $k$th binary treatment decision is, e.g., $a_k=1$ for treatment, $a_k=0$ for standard care. Patient information available at time $k$ and prior to $k$th treatment decision is denoted $\bx_k$. The covariate matrix containing patient history prior to the $k$th treatment decision is denoted $\boldsymbol{h}_{k}$; this history can include previous treatments $a_1, \ldots, a_{k-1}$. Finally, $\overline{\ba}_k = (a_1, a_2, \ldots, a_k)$ is the vector of the first $k$ treatment decisions, and $\underline{\ba}_k = (a_{k+1}, a_{k+2}, \ldots, a_K)$ is the vector of treatment decisions from stage $k+1$ onward.

The blip (or contrast) function is defined as the difference in expected potential outcome between patients who received treatment $a_k$ at stage $k$ and patients who received a reference treatment denoted, say $a_k=0$, with the same history and assuming they receive optimal treatment after $k$th stage: $$\gamma_{k}\left(\boldsymbol{h}_{k}, a_{k}\right)=\mathbb{E}\left[Y^{\overline{\ba}_{k}, \underline{\ba}_{k+1}^{opt}}-Y^{\overline{\ba}_{k-1},a_{k}=0,\underline{\ba}_{k+1}^{opt}} | \boldsymbol{H}_{k}=\bh_k\right].$$
The regret function \citep{Murphy} is the expected loss resulting from giving treatment $a_k$ at stage $k$ instead of the optimal treatment $a_k^{opt}$, assuming optimal treatment is received after $k$-th stage: $\mu_{k}\left(\boldsymbol{h}_{k}, a_{k}\right)=\mathbb{E}\left[Y^{\overline{\ba}_{k-1}, \underline{\ba}_{k}^{opt}}-Y^{\overline{\ba}_{k}, \underline{\ba}_{k+1}^{opt}} | \boldsymbol{H}_{k}=\boldsymbol{h}_{k}\right].$

The blip and regret functions correspond directly: $\mu_{k}\left(\boldsymbol{h}_{k}, a_{k}\right)=  \gamma_{k}\left(\boldsymbol{h}_{k}, a_{k}^{o p t}\right) -\gamma_{k}\left(\boldsymbol{h}_{k}, a_{k}\right)$. This can be leveraged to simplify some expressions in later sections. Finally, we decompose the expected mean outcome into two components: $\mathbb{E}\left[Y^{a} |\boldsymbol{H}=\bh;\boldsymbol{\beta},\boldsymbol{\psi}\right]=f\left(\boldsymbol{h}_{0}; \boldsymbol{\beta}\right)+\sum_{k=1}^{K}\gamma_{k}\left(\boldsymbol{h}_{k}, a_{k} ; \boldsymbol{\psi}_{k}\right),$ where $f\left(\boldsymbol{h}_{0}; \boldsymbol{\beta}\right)$ and $\gamma_{k}\left(\boldsymbol{h}_{k}, a_{k} ; \boldsymbol{\psi}_{k}\right)$ are the so-called treatment-free and blip models, respectively, and $\bh_0$ are baseline covariates. The function $f$, being free of any terms relating to the active treatment ($a_k=1$), is irrelevant for making decisions about optimal treatment selection. For instance, in a simple one-stage setting, we could assume that both $f$ and $\gamma$ are linear in form: $f(x;\bbeta)=\beta_0+\beta_1 x$ and $\gamma(x,a;\bpsi)=a(\psi_0+\psi_1x)$, and hence the estimated optimal treatment is $\widehat a^{opt}=I(\widehat\psi_0+\widehat\psi_1x >0)$ where $I(\cdot)$ is the indicator function.

\subsection{Dynamic weighted ordinary least squares}
Dynamic weighted ordinary least squares uses a sequential regression approach, similar to estimate the blip parameter $\boldsymbol{\psi}_{k}$ in the model for $\mathbb{E}\left[Y^{a} |\boldsymbol{H}=\bh;\boldsymbol{\beta},\boldsymbol{\psi}\right]$, achieving double robustness through weighting by a function of the propensity score \citep{ps}. The weights must satisfy $ \pi(\boldsymbol{x}) w(1, \boldsymbol{x})=(1-\pi(\boldsymbol{x})) w(0, \boldsymbol{x})$, where $\pi(\boldsymbol{x})$ is the propensity score and $w(a, \boldsymbol{x})$ is the weight for a subject with treatment $a$ and covariates $\boldsymbol{x}$. \citet{wallace2015doubly} suggested to use ``absolute value" weights of the form $ w(a, \bx) = |a - \mathbb{E}[A | \boldsymbol{X}=\boldsymbol{x}]|$, as these offered better efficiency than other alternatives considered, while yielding consistent estimators of blip parameters if either the treatment or treatment-free model is correctly specified. Another assumption required by dWOLS is that the treatment-free model must include the main effects for all covariates in the blip model (unlike G-estimation, which can use an intercept-only treatment-free model). Violation of this assumption, known as the strong heredity principle \citep{strong}, can lead to biased estimators of blip parameters.

\subsection{Penalized dWOLS}

We first introduce our approach in a one-stage setting with a continuous outcome, letting
\begin{align}
\bY  & = \beta_0\boldsymbol 1+ \psi_0 \boldsymbol{A}+\sum_{j=1}^p \bX_j \beta_j +  \sum_{j=1}^p\psi_{j} (\bA \circ \bX_j) + \boldsymbol{\varepsilon},  \label{eq:obj}
\end{align} where $\boldsymbol 1$ is the vector of $1$'s, $\bY \in \mathbb{R}^n$ is a continuous response measured on $n$ individuals, $\bX_j \in \mathbb{R}^{n}$ are the $j$-th covariates, $\bX_i \in \mathbb{R}^{p}$ are covariates of $i$-th individual, $\beta_j \in \mathbb{R}$ are the corresponding parameters for the main effects of covariates, $\psi_j \in \mathbb{R}$ are the blip parameters for $j=0, 1, \ldots, p$, $\bA$ is the binary treatment indicator, ``$\circ$" is the element wise vector multiplication, and $\boldsymbol{\varepsilon}$ is an error term. This model is a simplification of \citep{sahir}, which considers an additive interaction regression model. In this posited model, the treatment-free model is $\bbeta_0+ \sum_{j=1}^p \bX_j \beta_j$ and the blip model is $\psi_0 \boldsymbol{A}+ \sum_{j=1}^p\psi_{j} (\bA \circ \bX_j)$. To eliminate the intercept $\beta_0$, throughout this section, we center the response variable and each input variable in a weighted way, e.g., using $\bY-\frac{\sum_{i=1}^n w_i Y_i}{\sum_{i=1}^n w_i} $ instead of $\bY$ as the outcome.

For a continuous response we use the weighted squared-error loss: $$\mathcal{L}(\bY;\btheta) = \frac{1}{2n} \norm {\sqrt{\bW}\left(\bY  -\psi_0 \bA- \sum_{j=1}^p \bX_j \beta_j -  \sum_{j=1}^p   \psi_j (\bA \circ \bX_j)\right) }^2_2,$$ where $\btheta = (\beta_1, \ldots, \beta_p, \psi_0, \ldots, \psi_p)$, and $\bW=\diag \left\{w_1(a,\bx),w_2(a,\bx),\dots,w_n(a,\bx)\right\}$ is a \emph{known} $n\times n$ diagonal matrix with $w_i(a,\bx)$ the ``absolute value'' weight for the $i$th individual. Similar to LASSO, we consider the following objective function that includes the $\ell_1$ penalty for variable selection: \begin{equation}
Q\left(\btheta \right)= \mathcal{L}(\bY;\btheta) +\lambda(1-\alpha) \norm \bbeta _1 +\lambda \alpha \norm {\boldsymbol{\psi}} _1,
\label{eq:wobj}
\end{equation} where $\bbeta=(\beta_1,...,\beta_p)$ and $\boldsymbol{\psi}=(\psi_1,...,\psi_p)$, $\lambda>0$ and $\alpha \in (0,1)$ are tuning parameters, and the solution is given by
$
    \widehat{\btheta} = \argmin_{\btheta} Q(\btheta)
$. The parameter $\alpha$ controls the relative penalties for the main effects and the interaction effects. Other choices of the penalty term include the $\ell_2$ penalty, the elastic net \citep{net} and the SCAD penalty. The $\ell_2$ penalty yields ridge regression and hence cannot produce a sparse solution, and the $\ell_1$ penalty cannot handle highly correlated variables very well \citep{net}; the elastic net combines the $\ell_1$ and $\ell_2$ penalties, and thus can produce sparsity while offering good performance even when the features are highly correlated. The SCAD is a non-convex penalty that can produce sparse solutions and nearly unbiased estimators.

An issue with Equation \eqref{eq:wobj} is that since no constraint is placed on the structure of the model, it is possible that an estimated interaction term is nonzero while the corresponding main effects are zero, which violates the strong heredity assumption. To remedy this, our work is built on the strong heredity assumption, a constraint that is often used in practice when estimating interaction effects. Under the strong heredity assumption, an interaction term can be estimated to be non-zero if its corresponding main effects are estimated to be non-zero, whereas a non-zero main effect does not necessarily imply a non-zero interaction term. In DTR analysis, it is most common that there are more confounders than there are potential tailoring variables. Following \citep{choi}, we introduce a new set of parameters $\boldsymbol{\tau}=(\tau_1, \tau_2, ...\tau_p)$ and reparametrize the coefficients for the interaction terms $\psi_j$ as a function of $\tau_j$ and the main effect parameters $\beta_j$ and $\psi_0$: $\psi_j=\psi_0 \tau_j  \beta_j$. In this way, strong heredity can be met, and we consider the following model: $$
\mathcal{L}^*(\bY;\btheta) = \frac{1}{2n} \norm{ \sqrt{\bW}\left(\bY  -\psi_0 \bA- \sum_{j=1}^p \bX_j \beta_j -  \sum_{j=1}^p \underbrace{\psi_0 \tau_j  \beta_j}_{\textrm{$\psi_j$}} \left(\bA \circ \bX_j\right) \right)}_2^2,
$$ where now $\btheta = (\beta_1, \ldots, \beta_p, \psi_0, \tau_1, \ldots, \tau_p)$. This reparametrized model is nonlinear as it involves products of parameters, and the objective function is expressed as:
\begin{equation}
Q\left(\btheta \right)= \mathcal{L}^*(\bY;\btheta) +\lambda(1-\alpha) \norm \bbeta _1+\lambda \alpha \norm \btau _1.  
\label{eq:fobj}
\end{equation}

\subsection{Algorithm Details}
In this section, we describe a blockwise coordinate descent algorithm \citep{friedman2007pathwise} for fitting the weighted least-squares version of the model in Equation \eqref{eq:fobj}. ``Blockwise" means we breakdown the optimization problem into sub-problems, i.e., we fix the interaction terms $\btau$ and solve for the main effects $\psi_0$ and $\bbeta$ and vice versa. Following \citep{pathwise}, we fix the value for the tuning parameter $\alpha$ and minimize the objective function over a decreasing sequence of $\lambda$ values $(\lambda_{max} > \ldots > \lambda_{min})$. 

Denote the $n$-dimensional residual column vector $\bR = \bY-\widehat{\bY}$, where $\widehat{\bY}$ is the current fitted value of $\mathbb E(\bY)$ under the posited model. The subgradient equations are given by \begin{align}
\frac{\partial Q}{\partial \psi_0} & = -\frac{1}{n} (\bA + \sum_{j=1}^{p}\tau_j\beta_j \bA \circ \bX_j)^\top \bW \bR = 0 \label{eq:sub_bE}\\
\frac{\partial Q}{\partial \beta_j} & = -\frac{1}{n} \left(\bX_j + \tau_j \psi_0 \bA \circ \bX_j\right)^\top \bW \bR  + \lambda (1-\alpha) s_1 = \boldsymbol{0} \label{eq:sub_thetaj}\\
\frac{\partial Q}{\partial \tau_j} & = -\frac{1}{n} \left(\psi_0\beta_j \bA \circ \bX_j\right)^\top \bW \bR  + \lambda \alpha  s_2 = 0 \label{eq:sub_tauj}
\end{align}
where $s_1$ and $s_2$ are subgradients of the $\ell_1$-norm, i.e., $s_1 \in \textrm{sign}\left(\beta_j\right)$ if $\beta_j \neq 0$, $s_1 \in
[-1, 1] $ if $\beta_j=0$; $s_2 \in \textrm{sign}\left(\tau_j\right)$ if $\tau_j \neq 0$, $s_2 \in [-1, 1] $ if $\tau_j=0$.

Define the partial residuals, without the $j$th predictor for $j=1, \ldots, p$, as
\[\bR_{(-j)} = \bY -  \sum_{\ell \neq j} \bX_\ell \beta_\ell - \psi_0 \bA - \sum_{\ell\neq j} \tau_{\ell}  \psi_0 \beta_\ell \left(\bA \circ \bX_\ell\right) , \]
the partial residual without A as $\bR_{(-A)} = \bY - \sum_{j=1}^p \bX_j \beta_j$ and the partial residual without the $j$th interaction for $j=1, \ldots, p$, as
\[\bR_{(-jA)} = \bY  - \sum_{j=1}^p \bX_j \beta_j - \psi_0 \bA - \sum_{\ell\neq j} \tau_{\ell}  \psi_0 \beta_\ell \left(\bA \circ \bX_\ell \right). \]

From the subgradient Equations~\eqref{eq:sub_bE}--\eqref{eq:sub_tauj} we see that
$$ \widehat{\psi}_0  =\frac{ \left(\bA + \sum_{j=1}^{p}\tau_j \beta_j \left(\bA \circ \bX_j \right)\right)^\top \bW \bR_{(-A)} }{\left(\bA + \sum_{j=1}^{p}\tau_j \beta_j \left(\bA \circ \bX_j \right)\right)^\top \bW \left(\bA + \sum_{j=1}^{p}\tau_j\beta_j \left(\bA \circ \bX_j \right)\right)} $$

$$ \widehat\beta_j = \frac{S \left(\left(\bX_j + \tau_j \psi_0 \left(\bA \circ \bX_j \right)\right)^\top \bW \bR_{-j}, n \cdot \lambda (1-\alpha)\right)}{\left(\bX_j + \tau_j \psi_0 \left(\bA \circ \bX_j \right)\right)^\top \bW\left(\bX_j + \tau_j \psi_0 \left(\bA \circ \bX_j \right)\right)} $$

$$ \widehat\tau_j  = \frac{S \left(\left(\psi_0 \beta_j \left(\bA \circ \bX_j \right)\right)^\top  \bW \bR_{(-jA)}, n \cdot \lambda \alpha\right)}{\left(\psi_0 \beta_j \left(\bA \circ \bX_j \right) \right)^\top \bW \left(\psi_0 \beta_j \left(\bA \circ \bX_j \right)\right)} $$
where $S(x,u)$ is the soft-thresholding operator defined as $S(x,u)=\textrm{sign}(x) (\abs{x} - u)_+$ ($x_+$ is the maximum value of $x$ and $0$).

The strong heredity assumption means that finding the $\lambda$ which shrinks all coefficients to 0, is reduced to finding the smallest $\lambda$ such that all {\em main effect} coefficients are shrunk to 0. From the subgradient Equation~\eqref{eq:sub_thetaj}, we see that $\beta_j = 0$ is a solution if
$$
\abs{\frac{1}{n} \left(\bX_j + \tau_j \psi_0 \left(\bA \circ \bX_j \right)\right)^\top \bR_{(-j)}} \leq \lambda (1-\alpha).
$$
From the subgradient Equation~\eqref{eq:sub_tauj}, we see that $\tau_j = 0$ is a solution if
$$
\abs{\frac{1}{n} \left(\psi_0  \left(\bA \circ \bX_j \right)\beta_j\right)^\top \bR_{(-jA)}} \leq \lambda \alpha.
$$
Thus the strong heredity assumption implies that the parameter vector $(\beta_1, \ldots, \beta_p, \psi_1, \\ \ldots, \psi_p)$ will be entirely equal to $\boldsymbol{0}$ if $(\beta_1, \ldots, \beta_p) = \boldsymbol{0}$. Therefore, the smallest value of $\lambda$ for which the entire parameter vector reduces to
$
\lambda_{max}=\frac{1}{n(1-\alpha)} \max_j \left\lbrace \abs{\left(\bX_j\right)^\top \bR_{(-j)}} \right\rbrace.
$ The computational algorithm to fit all the parameters in a sequence of loops is further detailed in the Supplementary Material (Algorithm 1).

\subsection{Multiple Intervals Estimation}

Knowing how to estimate the blip parameters in a one-stage setting, we now describe how the pdWOLS approach works in a $K$-stages setting. Starting from the last stage, the estimation procedure is applied to the $K$-th stage observed outcome $\by_K$, treatment $\ba_K$, and covariates $\bx_K$. The estimated blip parameters are obtained by maximizing the objective function in Equation \eqref{eq:fobj} and the estimated rules $\widehat a_K^{opt}=I(\widehat\psi_{0K}+\bx_K \boldsymbol{\widehat\psi_K}>0)$, where $I$ is the indicator function. The ($K$-1$)$-th stage outcome is based on ``optimal responses", that is, the estimation procedure is applied to the pseudo-outcome $\widetilde \by_{k-1}=\by_K+\mu_K(\bx_K,a_K;\boldsymbol{\widehat\psi_K})$, treatment $\ba_{K-1}$ and covariates $\bx_{K-1}$, where $\mu_K(\bx_K,a_K;\boldsymbol{\widehat\psi_K})=\gamma_{K}(\bx_K,\widehat a_K^{opt};\boldsymbol{\widehat\psi_K})-\gamma_K(\bx_K,a_K;\boldsymbol{\widehat\psi_K})$ is the regret function at stage $K$. The pseudo-outcome, $\widetilde \by_{K-1}$, is optimal since the regret is added to the observed outcome $\by_K$. The same procedure continues, recursively working backwards, until stage $1$ estimation, such that the blip parameters across all the stages are obtained and all treatment decisions can be made.

\subsection{Asymptotic Properties of the pdWOLS estimator}

We now show that when the number of predictors, $p$, is fixed and the sample size $n$ approaches infinity, the pdWOLS estimator has both the double robustness and oracle properties \citep{SCAD} under several assumptions. Following the adaptive LASSO \citep{adaptive}, we add adaptive weights (or penalty factors) to the objective function~\eqref{eq:fobj} to obtain \begin{equation}
    \mathcal{L^*}(\bY;\btheta) +\lambda(1-\alpha)\sum_{j=1}^p w_j^{main} |\beta_j|+\lambda \alpha \sum_{j=1}^p w_j^{int} |\tau _j|,
\label{eq:wobj2}
\end{equation}
where $w_j^{main}$ and $w_j^{int}$ are adaptive weights of main effect and interaction terms respectively, in this way, the coefficients are not forced to be equally penalized in the $\ell_1 $ penalty. For instance, we can choose $w_j^{main}=\abs{\widehat\beta_j^{wls}}^{-1}$ and $w_j^{int}=\abs{\frac{\widehat\beta_j^{wls}\widehat\psi_0^{wls}}{\widehat\psi_j^{wls}}}$ for penalty factors, where $\widehat \beta_j^{wls}$ and $\widehat \psi_j^{wls}$ are unpenalized weighted least square estimates of the pdWOLS model. As $n$ goes to infinity, the weights corresponding to unimportant variables go to infinity, which puts a large penalty on those variables, and the weights corresponding to important variables converge to a finite constant. Thus, small coefficients are removed, and large coefficients are unbiasedly estimated. Without loss of generality, we can rewrite Equation \eqref{eq:wobj2} as $\mathcal{L}^*(\bY;\btheta) +\sum_{j=1}^p \lambda_j^{\beta} |\beta_j|+ \sum_{j=1}^p \lambda_j^{\tau}  |\tau _j|,$ where $\lambda_j^{\beta}=\lambda(1-\alpha)w_j^{main}$ and $\lambda_j^{\tau} =\lambda\alpha w_j^{int}$.

We assume that the true model follows the strong heredity assumption described above and regularity conditions detailed in the Supplemental Material hold. Note that the regularity conditions of pdWOLS are for quasi-likelihood since the loss function contains data-dependent weights and the treatment-free model may be misspecified. We describe the asymptotic properties of pdWOLS in the following theorems; proofs are given in the Supplemental Material. Assume that the observations $\bV_i, i=1, \dots, n$ are independent and identically distributed with probability density $g(\bV)$ with respect to a measure $\nu$. Denote the negative quasi-log-likelihood as $L_n^*(\bV;\btheta) = -\sum_{i=1}^n log\, h(\bV_i,\btheta)$ (i.e., the dWOLS loss function), where $h$ is the posited family of densities. Let $\btheta^*$ be the underlying true parameters, and $\btheta_*$ the minimizer of the Kullback–Leibler divergence between $h$ and $g$ (i.e., $\btheta_*$ is the closest point to $\btheta^*$ in the posited family of densities). Define $B_1$ as the indices of non-zero components for main effects and $B_2$ as the indices of non-zero components for interaction terms such that $$ B_1=\{j:\beta_{*j}\neq0\}, B_2=\{j+p+1:\tau_{*j}\neq0\}, B=B_1\cup B_2,$$ where we define $\btau_*$ in a way such that $\tau_{*j}=\frac{\psi_{*j}}{\psi_{*0}\beta_{*j}}$ if $\beta_{*j} \neq 0$ and 0 otherwise, since we assume the strong heredity property holds. Let $na_n$ be the maximum value of the tuning parameters $(\lambda^{\beta}, \lambda^{\tau})$ such that the corresponding coefficients are non-zero and $nb_n$ be the minimum value of the tuning parameters such that the corresponding coefficients are zero. For $\lambda^{\tau}$ we only consider the index $m$ such that $\beta_{*m}\neq0$ and $\psi_{*m}=0$ (i.e., $m \in B_1$): $$a_n= \frac{1}{n} max \{\lambda_j^{\beta}, \lambda_m^{\tau}: j\in B_1, m+p+1\in B_2\}$$ $$b_n= \frac{1}{n} min\{\lambda_j^{\beta},  \lambda_m^{\tau}:j\in B_1^c, m+p+1 \in B_2^c \text{  such that } \beta_{*m}\neq0\}.$$

\begin{theorem}
Correct Sparsity: Assume that $\sqrt n a_n=O(1)$ and $\sqrt{n} b_n\to \infty$, then there exists a local minimizer $\widehat\btheta_n$ of Equation \eqref{eq:wobj2} such that $\norm {\widehat\btheta_n-\btheta_*}=O_p(n^{-\frac{1}{2}}+a_n)$. Moreover, we have $P(\widehat\btheta_{B^c}=0)\to 1.$
\end{theorem}

\begin{theorem}
Asymptotic Normality: Assume that $\sqrt{n}a_n\to 0$ and $\sqrt{n} b_n\to \infty$, then $$\sqrt{n}(\widehat\btheta_B-\btheta_{*B})\to_d N\left(0,\boldsymbol{J}^{-1}(\btheta_{*B})\boldsymbol{I}(\btheta_{*B})\boldsymbol{J}^{-1}(\btheta_{*B})\right)$$ where $\boldsymbol{J} (\btheta)=-  E_{\btheta}\left[\frac{\partial^2 log \; h(\bV;\btheta)}{\partial \btheta \partial \btheta^T}\right]$ and $\boldsymbol I (\btheta)=  E_{\btheta}\left[\left(\frac{\partial log \;h(\bV;\btheta)}{\partial \btheta}\right) \left(\frac{\partial log \; h(\bV;\btheta)}{\partial \btheta}\right)^T\right]$.
\end{theorem}

\begin{remark}
Oracle properties of $\widehat \btheta_n$ are established such that the estimator converges to some population parameter instead of the underlying true parameter $\btheta^*$. Also, the asymptotic covariance matrix no longer equals the inverse of the Fisher's information matrix. If the treatment-free model is correctly specified, then $\widehat\btheta_n$ will converge to $\btheta^*$. To mimic the oracle, we further assume that all the observational weights are $1$ (e.g., as in a randomized study).
\end{remark}

\begin{corollary}
Double Robustness: Assume that the blip function is correctly specified and SUTVA and ignorability described in Section 2.1 hold, then the resulting blip parameter estimators of pdWOLS are doubly-robust; the estimators are consistent (i.e., $\bpsi_*=\bpsi^*$) if either the treatment model or the treatment-free model is correct. Note that correct specification of the blip model permits \textit{over-specification} - that is, the true blip model may be contained within the analyst-specified model. From Theorems 1 and 2, pdWOLS has the same performance as dWOLS, and hence it has the double robustness property. 
\end{corollary}

\begin{remark}
There are no consistency guarantees for the first-stage estimator if an important confounder is missing in the second-stage model, as this violates an assumption at the second stage such that the estimator of second-stage parameters (subsequently plugged into the first-stage estimating function) may be biased. However, if estimation at the second stage is consistent (no unmeasured confounding, at least one of the nuisance models correct, etc), then double-robustness at the first stage can  be assured under key assumptions.
\end{remark}

\section{Simulation Studies}

In this section, we first illustrate the double robustness of pdWOLS and compare its performance to competing approaches through a number of simulations; then we implement the proposed method in a high  dimensional setting where $p>n$. Lastly, we present simulation results for a two-stage setting. The tuning parameter $\alpha$ was set to 0.5 for all simulations, and $\lambda$ was selected using four-fold cross-validation to reduce the computational burden. 

In addition to assuming that there are no unmeasured confounders, we assume that the number of confounders is relatively small, so that the propensity score model can be fitted using logistic regression with the entire vector $\bX$. The propensity score is used to ensure balance between treatment groups. If model misspecification is a concern, one can use data-adaptive techniques, however, care must be taken in using data-adaptive approaches to estimating the propensity score to avoid the risk of selecting instruments, i.e., variables that only predict treatment \citep{susan}. To consider a general framework, main effects are penalized in Equation (\ref{eq:fobj}). However, in a low dimensional setting, we may want to retain all available covariates in the outcome model to ensure no weak confounders are erroneously omitted. In such cases, we can choose to not penalize the main effects, setting the corresponding penalty factors in Equation (\ref{eq:wobj2}) to zero.


\subsection{Competing Methods}

We compare the variable selection results, error rate (in terms of the estimated rules as compared to the true optimal treatment), and out-of-sample value (i.e., expected outcome) under the estimated rules of pdWOLS with Q-learning combined with LASSO \citep{LASSOq} and penalized A-Learning (PAL) \citep{al}. Q-learning is a sequential regression approach to DTR estimation; relying only on outcome models; it is not doubly robust. PAL first estimates the treatment-free and propensity score models, then uses the Dantzig selector \citep{Dan} to penalize the estimating equations of A-learning: $\widehat\bpsi=arg min_{\bpsi}\: \norm{\bpsi}_1$ subject to $\norm{\bX^T diag (\bA- \boldsymbol{\widehat\pi})(\bY-f(\bx;\widehat \bbeta)-\gamma(\bx,a;\bpsi))}_{\infty} \leq n\lambda_{pal}$, where $\lambda_{pal}$ is the tuning parameter and $\boldsymbol{\widehat\pi}$ is the estimated propensity score.

LASSO was implemented using the R package glmnet \citep{pathwise} with $\lambda_{LASSO}$ selected via four-fold cross-validation. PAL was implemented using the R package ITRSelect \citep{al} with the tuning parameter $\lambda_{pal}$ selected via the Bayesian Information Criteria (BIC) \citep{BIC}. The main effect of treatment $A$ is not penalized in any of the three methods. 
We also present unpenalized estimates of the blip parameters from a two-step approach: that is, after variable selection, the blip parameters are re-calculated by solving the unpenalized weighted least squares via Q-learning, dWOLS, and A-learning with the selected variables, which we term 
\textit{refitted} procedure.

\subsection{Experiments Examining Double Robustness Property}

We begin with a simple one-stage example with the following data generation procedure:

Step 1: Generate 10 covariates ($\bX_1-\bX_{10}$) where $\bX$ are multivariate normal with zero mean, unit variance, and correlation $Corr(X_j,X_k)=0.25^{|j-k|}$ for $j, k= 1, 2,  \dots, 10$.

Step 2: Generate treatment according to the model: $$P(A=1|X_1,X_2)=\cfrac{exp(1+x_1+x_2)}{1+exp(1+x_1+x_2)}.$$   

Step 3: Set the blip function, and hence the optimal treatment strategy, to depend only on $X_1$: $\gamma(x,a;\bpsi)=a(\psi_0+\psi_1x_1)$ for $\psi_0=1, \psi_1=-1.5$.

Step 4: Set the treatment-free model to $f(\bx;\bbeta)=0.5-0.6e^{x_1}-2x_1-2x_2$.

Step 5: Generate the outcome $Y \sim  N(f(\bx;\bbeta)+\gamma(\bx,a;\bpsi),1).$ 

We apply estimation and variable selection approaches with a variety of sample sizes (100, 500, and 2000) in four scenarios, where neither, one, or both of the treatment and treatment-free models is correctly specified. Specifically, the scenarios are: \textit{Scenario 1 (neither treatment nor treatment-free is correct):} Regress $\bY$ on ($\boldsymbol{1},\bX,\bA, \bA\bX$), and set all observational weights to 1 (similar to assuming a null propensity score model). As this scenario fails to meet the assumptions of correct model specification, consistency is not assured for any approach. \textit{Scenario 2 (treatment correct, treatment-free incorrect):} Regress $\bY$ on ($\boldsymbol{1},\bX,\bA, \bA\bX$), but fit a correctly specified propensity score model whose parameters are estimated via logistic regression. \textit{Scenario 3 (treatment incorrect, treatment-free correct):} Regress $\bY$ on ($\boldsymbol{1},e^{\bX_1},\bX,\bA,\bA e^{\bX_1}, \bA\bX$), so that the treatment-free model is correctly specified but - as in scenario 1 - set all observational weights to 1.  \textit{Scenario 4 (both treatment and treatment-free are correct):} Regress $\bY$ on ($\boldsymbol{1},e^{\bX_1},\bX,\bA,\bA e^{\bX_1}, \bA\bX$), and estimate the parameters using a correctly specified propensity score.

Since Q-learning does not incorporate any propensity score adjustments, scenarios 1 and 2 yield identical estimates, as do scenarios 3 and scenario 4. All the three methods have the same treatment-free models and the same blip functions to be estimated in the four scenarios. Across all scenarios where at least one nuisance model was correctly specified, refitted estimators performed better than their penalized counterparts in terms of bias (see Figure S1 in the Supplementary Material). When at least one of the treatment or treatment-free models was correctly specified, the blip parameter estimators were consistent for refitted pdWOLS. When the treatment-free model was correct (Scenarios 3 and 4), the refitted  Q-learning (LASSO) estimators were consistent, as expected. Surprisingly, PAL failed when the treatment model was incorrect (Scenario 3). This result was not anticipated since PAL is a double robust method, although previous simulations have not considered its performance in terms of parameter estimates \citep{al}.

The variable selection results for optimal treatment decisions are presented in Table 1. In Scenarios 2-4, the important tailoring variable was correctly selected by both pdWOLS and Q-learning (LASSO). PAL failed in scenario 3. However, the false positive rates of pdWOLS and Q-learning (LASSO) were higher than that of PAL in all scenarios: for example, in Scenario 3, both LASSO and pdWOLS falsely selected the variable $Ae^{X_1}$ $72\%$ of the time. 

Table 1 also summarizes the error rates (i.e., $\frac{1}{n}\sum_{i=1}^n I(a_i^{opt}\neq \widehat a_i)$) of the estimated optimal treatment regimes for treatment decision making and value functions. The average value function and the error rates were computed over a testing set of size 10,000 (i.e., a dataset generated according to the process described above in all respects except that treatment was allocated according to the estimated rule). Both the error rate and the value of pdWOLS and Q-learning with LASSO were very close; pdWOLS outperformed other methods in Scenario 2, while Q-learning with LASSO had the best performance in Scenarios 3 and 4. The performance of the refitted versions of pdWOLS and Q-learning were similar; the performance of PAL was uniformly worse than the other methods performed without refitting, however refitting PAL substantially improved its performance.

\begin{table}[H]
\caption{Variable selection rate (\%) of the blip parameters, error rate (ER, \%) and value function over a testing set of size 10,000 under the estimated decision rules using pdWOLS, Q-learning with LASSO (QL), PAL and their refitted versions ($n=500$, $400$ simulations). The main effect of treatment is not penalized (and hence is always selected).}
\centering
\begin{threeparttable}
\resizebox{\textwidth}{!}{%
\begin{tabular}[t]{lrrrrrrrrrrrr}
\toprule
\multicolumn{1}{c}{} & \multicolumn{3}{c}{Scenario 2} & \multicolumn{3}{c}{Scenario 3} & \multicolumn{3}{c}{Scenario 4} \\
\cmidrule(l{3pt}r{3pt}){2-4} \cmidrule(l{3pt}r{3pt}){5-7} \cmidrule(l{3pt}r{3pt}){8-10} \cmidrule(l{3pt}r{3pt}){11-13}
& pdWOLS & QL  & PAL & pdWOLS & QL & PAL & pdWOLS & QL & PAL\\
\midrule
$A e^{X_1}$  & - & - & - & 72 & 14 & 72 & 42 & 14 & 0\\
$AX_1^*$  & 100 & 100 & 99 & 100 & 100 & 33 & 100 & 100 & 100\\
$AX_2$ & 53 & 51 & 2 & 73 & 44 & 3 & 52 & 44 & 1\\
$AX_3$ & 2 & 26 & 2 & 6 & 23 & 0 & 2 & 23 & 1\\
$AX_4$  & 4 & 28 & 2 & 5 & 24 & 1 & 3 & 24 & 1\\
$AX_5$  & 4 & 29 & 4 & 4 & 26 & 2 & 2 & 26 & 2\\
$AX_6$& 2 & 26 & 2 & 4 & 21 & 1 & 1 & 21 & 1\\
$AX_7$ & 3 & 25 & 2 & 5 & 22 & 1 & 2 & 22 & 0\\
$AX_8$ & 3 & 27 & 3 & 6 & 23 & 1 & 2 & 23 & 0\\
$AX_9$ & 2 & 27 & 3 & 6 & 24 & 1 & 2 & 24 & 1\\
$AX_{10}$ & 2 & 28 & 2 & 5 & 22 & 1 & 1 & 22 & 1\\
\bottomrule

ER  & 3.9 &9.8 & 22.9 & 5.5 & 3.4 & 12.0 & 4.2 & 3.4 & 23.4\\
ER (Refitted) & 4.5 &8.5 & 4.9 & 3.4 & 3.6 & 8.7 & 3.6 & 3.6 & 3.8\\
Value & 0.6 &0.6 & 0.5 & 0.6 &0.7 & 0.6&0.6 &0.7 & 0.5\\
Value (Refitted)& 0.6 &0.6 & 0.6 & 0.6 &0.7 & 0.6&0.7 &0.7 & 0.6\\
\bottomrule
\end{tabular}
}
\begin{tablenotes}
\item [*] Term with a non-zero coefficient in the data-generating model
\item [] Note that $A e^{X_1}$ was not included in the blip model for scenario 2
\end{tablenotes}
\end{threeparttable}
\end{table}

\subsection{Simulations Evaluating Performance in a High-dimensional Setting}

Here we present the performance of the new procedure in a high dimensional setting with $p=400$ and $n=200$. The data generation procedure is the same as in Section 3.2, except that we now set $P(A=1)$ to 0.5 for everyone such that no confounding is present. The blip function is $\gamma(\bx,a;\bpsi)=a(1-1.5x_1)$ where $\psi_0=1, \psi_1=-1.5$ and the treatment-free model is $f(\bx;\bbeta)=0.5-0.6e^{x_1}-2x_1-2x_2$. We regress $\bY$ on $(\boldsymbol{1}, \bX, \bA, \bA\bX)$ where the treatment-free model is misspecified. 

Figure 1 summarizes the blip parameter estimates in the high dimensional setting. Like before, for all the methods, refitted estimators improved the performance of their penalized counterparts. For $\psi_0$, Q-learning with LASSO and its refitted estimator had the smallest bias; as for $\psi_1$, pdWOLS and its refitted version had the smallest bias.

\begin{figure}[H]
    \centering
    \centerline{\includegraphics[width=\textwidth]{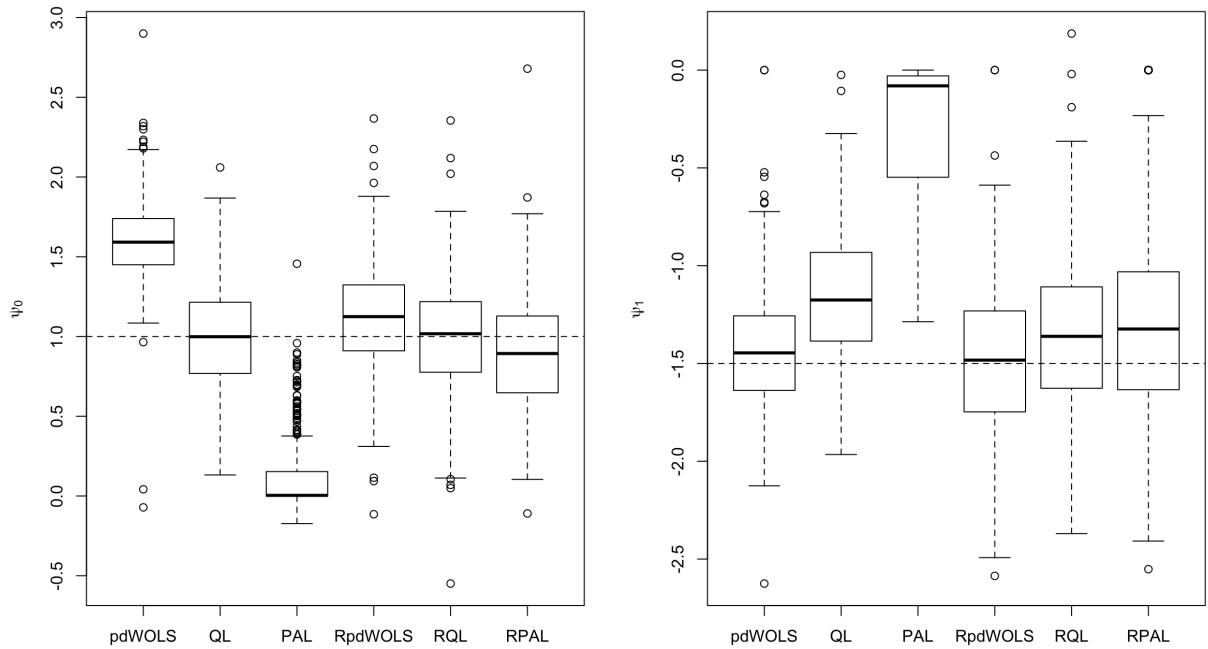}}
    \caption{Estimates of blip parameters using pdWOLS, Q-learning (LASSO), PAL and their refitted versions with sample size 200 (400 simulations) in a high dimensional ($p=400$) setting. The true value is represented by the dotted line.}
\end{figure}

Table 2 shows false negative rates (the proportion of times a method wrongly removed a truly important variable), false positive rates (the proportion of times a method wrongly included a non-important variable), error rates, and the value under the estimated rules of the three methods. The average value function and the error rates were computed over a testing set of size 10,000. Q-learning with LASSO achieved a zero false negative rate; pdWOLS and refitted pdWOLS had the lowest false positive rate, error rate, and the highest value, which indicates favorable performance of the newly proposed method. However, unlike before, even the refitted PAL estimator had a smaller bias than the PAL estimator; refitted PAL did not improve the performance of PAL with respect to value and error rate, which shows that smaller bias in estimation of blip parameters does not necessarily translate into a better performance of the estimated regime.

\begin{table}[H]
\caption{False negative (FN, \%) rate and false positive (FP, \%) rate of variable selection results of the blip parameters, error rate (ER, \%) and value using pdWOLS, Q-learning with LASSO (QL), PAL and their refitted versions with sample size 200 (400 simulations) in a high dimensional ($p=400$) setting. The main effect of treatment is not penalized (and hence is always selected).}
\centering
\begin{threeparttable}
\begin{tabular}[t]{lrrrr}
\toprule
  & FN & FP  & ER &  Value\\
\midrule
pdWOLS & 0.3 & 0.2 & 12.8  & 0.7\\
QL (LASSO) & 0.0 & 1.4 & 11.3  & 0.7\\
PAL & 2.6 & 0.4 & 24.6  & 0.6\\

RpdWOLS & 0.3 & 0.2 & 9.9 &  0.7\\
RQL (LASSO) & 0.0 & 1.4 & 16.8  & 0.6\\
RPAL & 2.6 & 0.4 & 25.1 &  0.5\\
\bottomrule
\end{tabular}
\end{threeparttable}
\end{table}

\subsection{Simulations Evaluating Performance in Multi-stage Setting}

In this subsection, we demonstrate the performance of the proposed pdWOLS approach when treatment decisions are made at multiple stages. We consider two different data generation procedures in order to follow previous literature. Setting 1, in which the true treatment-free model does not have an analytical closed-form (misspecified treatment-free model) is presented here. Setting 2, in which the treatment-free models can be computed analytically, is available in the Supplemental Material.

We follow the data generation procedure in \citep{wallace2015doubly} with a sample size of 1000:

Step 1: Generate 10 covariates at stage 1: $X_{j1}\sim N(0,1)$ for $j=1, 2,...10$.

Step 2: Generate treatment at stage $k$ according to $$P(A_k=1|X_{1k},X_{2k})=\frac{exp(x_{1k}-x_{2k})}{1+exp(x_{1k}-x_{2k})},$$ for $k=1, 2$.

Step 3: Generate covariates at stage 2, such that $X_{12}\sim N (0.5A_1+0.8X_{11},1)$ and $X_{j2}\sim N(0.8X_{j1},1)$, for $j=2, 3,...10$.

Step 4: Set the blip functions to be  $\gamma_1(x_1,a_1;\bpsi_1)=a_1(0.8-2x_{11})$ and   
$\gamma_2(x_2,a_2;\bpsi_2)=a_2(1-1.5x_{12})$, so that $\psi_{01}=0.8$, $\psi_{11}=-2$, $\psi_{02}=1$ and $\psi_{12}=-1.5$.

Step 5: Generate the outcome under optimal treatment according to $y^{opt}=0.5+2x_{11}+2x_{12}$. The observed outcome is generated such that $Y\sim N(y^{opt}-\mu_1-\mu_2,1)$ where $\mu_1$ and $\mu_2$ are regret function at stages 1 and 2, defined through the blip functions in step 4.

Recall, that a backward recursive approach can be used to make the treatment decision. Starting from the last stage, the estimation procedure is applied to the observed outcome $\by$. The estimated blip parameters and the estimated rules, $\widehat a_2^{opt}$, are obtained. Estimation then proceeds to stage 1, where again the estimation procedure is applied to a pseudo-outcome which represents the expected effect of the observed stage 2 treatment with the optimal stage 2 treatment. In pdWOLS, the pseudo-outcome is $\widetilde y_1=y+\gamma_2(\bx_2,\widehat a_2^{opt};\widehat \bpsi_2)-\gamma_2(\bx_2,a_2;\widehat \bpsi_2)$, where as for Q-learning with LASSO, the pseudo-outcome is $\widetilde y_1^{Q}=f(\bx_2;\widehat\bbeta_2)+\gamma_2(\bx_2,\widehat a_2^{opt};\widehat \bpsi_2)$.

In this setting, the treatment free model in the second stage of estimation aims to represent $y^{opt}-\mu_1-a_2^{opt}(1-1.5x_{12})$ which depends on $a_2^{opt}$, which in turn is a function of second stage parameters $\bpsi_2$ and covariate $x_2$. The treatment free model in this setting cannot be computed analytically. We nevertheless assumed that the treatment-free models were linear in the covariates measured at their respective stages, and thus in these simulations, it is always the case that the treatment-free models were misspecified. For those methods relying on a propensity score, the treatment models were fit using correctly-specified logistic regression models at each stage using all covariates measured at that stage.

Figure 2 summarizes the estimates of blip parameters using the three methods in the two-stage Setting 1. As expected, pdWOLS and PAL work when at least one of the treatment or treatment-free models is correctly specified (in this case, the treatment model is correctly specified), and Q-learning with LASSO failed, since the treatment free model at both stages are misspecified. For pdWOLS and PAL, refitted estimators were nearly unbiased, and they performed better than their penalized counterparts. At stage 1, the bias of PAL estimators decreased to almost zero after refitting. Thus, PAL exhibits excellent performance in variable selection but requires the additional step of refitting for accurate estimation. Unlike PAL, pdWOLS can have small bias even without the refitting procedure. 

\begin{figure}[H]
    \centering
    \includegraphics[width=\textwidth]{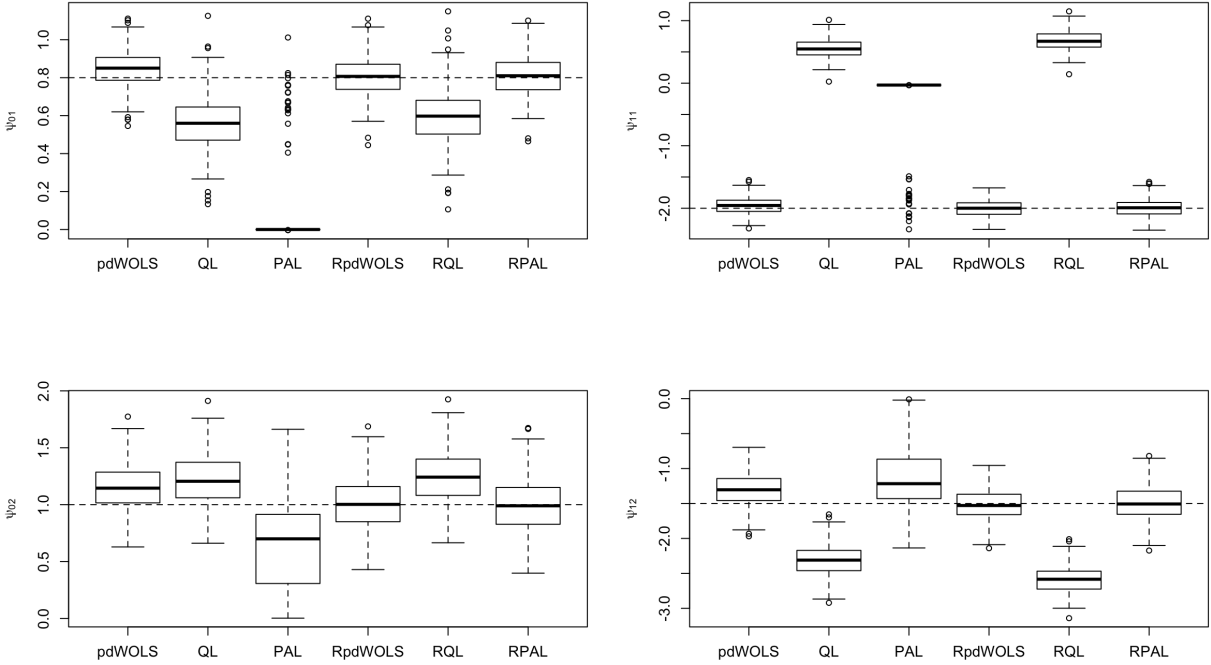}
    \caption{Estimates of blip parameters using pdWOLS, Q-learning with LASSO (QL), PAL and their refitted versions with sample size 1000 (400 simulations) in two-stage Setting 1. The true value is represented by the dotted line.}
\end{figure}

Table 3 presents the variable selection results for optimal treatment decisions. The important tailoring variables were selected by all methods at both stages. At stage 2, the false positive rate of pdWOLS was much smaller than other two methods. For instance, the selection frequency of $AX_2-AX_{10}$ were all less than $5\%$. Note that at stage 1, because the pseudo-outcomes were different for refitted version and their penalized counterparts, the variables selected by the procedures may differ between penalized and unpenalized implementations.

\begin{table}[H]
\caption{Variable selection rate (\%) of the blip parameters using pdWOLS, Q-learning with LASSO (QL), PAL and their refitted versions with sample size 1000 (400 simulations) in two-stage Setting 1. The main effect of treatment is not penalized (and hence is always selected).}
\centering
\begin{threeparttable}
\begin{tabular}[t]{lrrrrrrrrr}
\toprule
\multicolumn{1}{c}{} & \multicolumn{6}{c}{Stage 1} & \multicolumn{3}{c}{Stage 2} \\
\cmidrule(l{3pt}r{3pt}){2-7} \cmidrule(l{3pt}r{3pt}){8-10}
&pdWOLS& QL&PAL &RpdWOLS& RQL&RPAL&pdWOLS & QL &  PAL\\
\midrule
$AX_1$ *& 100 & 100 & 100 & 100 & 100 & 100 & 100 & 100 & 100\\
$AX_2$ & 49 & 32 & 1 & 45 & 33 & 2 & 22 & 44 & 33\\
$AX_3$ & 4 & 28 & 2 & 2 & 34 & 2 & 2 & 41 & 38\\
$AX_4$ & 3 & 30 & 0 & 2 & 34 & 2 & 3 & 45 & 37\\

$AX_5$ & 3 & 25 & 1 & 2 & 29 & 2 & 2 & 40 & 40\\
$AX_6$& 4 & 25 & 1 & 2 & 29 & 1 & 3 & 40 & 40\\
$AX_7$ & 4 & 27 & 0 & 2 & 33 & 2 & 3 & 40 & 38\\
$AX_8$ & 4 & 28 & 0 & 2 & 30 & 1 & 2 & 42 & 38\\
$AX_9$ & 4 & 26 & 1 & 2 & 32 & 4 & 2 & 41 & 36\\

$AX_{10}$ & 3 & 29 & 2 & 2 & 32 & 2 & 2 & 44 & 38\\
\bottomrule
\end{tabular}
\begin{tablenotes}
\item [*] Term with a non-zero coefficient in the data-generating model
\end{tablenotes}
\end{threeparttable}
\end{table}

Table 4 summarizes the error rates of the estimated optimal treatment decisions and value functions, computed over a testing set of size 10,000. As before, refitted methods had lower error rate and higher value functions than their penalized counterparts. Penalized dynamic ordinary least squares outperformed other methods at both stages with respect to the error rate and value function; refitting greatly improved the performance of PAL.

\begin{table}[H]

\caption{Error Rate (\%) and value function using pdWOLS, Q-learning  with LASSO (QL), PAL and their refitted versions with sample size 1000 (400 simulations) in two-stage Setting 1. The total error rate (TER, \%) in the estimated optimal treatment across both stages as well as the stage-wise error rates are shown. }

\centering
\begin{threeparttable}
\begin{tabular}[t]{lrrrr}
\toprule
  & TER & ER (Stage 1) & ER (Stage 2) & Value\\
\midrule
pdWOLS & 9.2 & 2.2 & 7.2 & 0.4\\
QL (LASSO) & 52.8 & 50.4 & 4.6 & -0.6\\
PAL & 22.4 & 14.5 & 9.8 & 0.4\\

RpdWOLS & 6.5 & 2.0 & 4.6 & 0.5\\
RQL (LASSO) & 58.0 & 54.8 & 6.1 & -0.7\\
RPAL & 11.3 & 2.1 & 9.5 & 0.4\\
\bottomrule
\end{tabular}
\end{threeparttable}
\end{table}

Additionally, we compared the choice of tuning parameter $\alpha$, in order to assess sensitivity of the results to this choice; we considered values of 0.2, 0.5 (as in the analyses above), and 0.8. The results are presented in the Supplemental Material (Figure S4, Tables S3 and S4). To briefly summarize, among all the $\alpha$'s, the bias and the variance of the estimators, the error rate and the estimated value were virtually identical. However, for variable selection, as $\alpha$ increased, the false positive rate decreased notably (See Table S3 in the Supplemental Material), as a larger $\alpha$ will put more penalty on the interaction terms. 

\section{Application to STAR*D Study}

In this section, we apply pdWOLS to STAR*D data \citep{stard} from the NIMH Data Archive, a multistage randomized trial that aimed to determine effective treatments for patients with major depressive disorder, where severity was measured using the Quick Inventory of Depressive Symptomatology (QIDS) score \citep{qids}. The study was divided into four levels (one of which had two sub-levels); patients had different treatments at each level would exit the study upon achieving remission. See the Supplemental Materials for details.

We follow \citet{modelg} and \citet{sb} to perform two-stage analysis based on the use of a selective serotonin reuptake inhibitor (SSRI), with negative QIDS score as the outcome. Three tailoring variables were considered: (1) the QIDS score measured at the beginning of each level (denoted by $q_k$ at stage $k$); (2) change in QIDS score divided by the time in the previous level (QIDS slope, denoted by $s_k$ at stage $k$); and (3) patient preference measured prior to receiving treatment, which is a binary variable (denoted by $p_k$ at stage $k$). We also generated $d$ iid noise variables at each stage: noise variables at stage 1 were generated using $X_{j1} \sim N(0,1) $ and at stage 2, $X_{j2} \sim N(log\abs{X_{j1}},1)$ for $j=1, 2, \dots, d$. We consider three scenarios for the analysis where $d=5, 10, 20$ respectively.

Logistic regression was used to estimate the treatment model adjusting for patient preference only, following the trial design, and weights $w=\abs{A-E(A|X)}$ were used in the analysis. As in \citet{modelg}, the treatment‐free models were linear in $(q_1, s_1, p_1)$ at stage 1 and  $ (a_1, q_2, s_2, p_2) $ at stage 2. Linear blip models with covariates $(q_1, s_1, p_1)$ at stage 1 and $(a_1,  q_2, s_2, p_2)$ at stage 2 were considered. Note in \citep{modelg}, $a_1$ and $p_2$ were not included in the blip models to avoid the multicollinearity; this is not necessary in pdWOLS, and hence our model specifications differ.

As in our simulations, the main effect of treatment was not penalized. In all three scenarios and both stages, pdWOLS returned the intercept‐only blip model, suggesting that the optimal treatments are treat with SSRI ($A_1=1$) and treat with a non-SSRI ($A_2=0$) at stage 1 and 2, respectively, for all patients. Penalized A-learning, in contrast, was sensitive to the number of noise variables: when $d=5$, PAL selected $a_j, q_j, s_j, p_j$ for both stages $j=1,2$. When $d=10$, PAL selected $a_2$ at stage $2$ and $a_1,q_1,s_1,p_1$ at stage $1$, and when $d=20$, PAL selected $a_2$ at stage $2$ and $a_1, p_1$ at stage $1$. \citet{sb} and \citet{modelg} found that no stage 2 blip covariates were statistically significant (consistent with pdWOLS), while at stage 1, they found only treatment preference was significant. 

The false positive rates of PAL at stage $2$ and $1$ were $100\%$, $40\%$ ($d=5$),  $10\%$, $50\%$ ($d=10$), and $10\%$, $45\%$ ($d=20$), respectively; for pdWOLS, the rate was 0\% for all $d$.

\section{Discussion}

In this article, we extended dWOLS to a penalized estimation framework for variable selection and estimating the optimal treatment regimes simultaneously. The proposed method inherits the double robustness property from dWOLS. Our simulations indicated that pdWOLS compares favorably with other variable selection approaches in the context of DTRs. 

Our method automatically enforces strong heredity through a simple reparametrization, which guarantees an assumption required by dWOLS. The idea of reparametrization is simple, however, one limitation is that the objective function is non-convex. Hence, it may be of interest, in future work, to investigate approaches that use convex constraints to achieve strong heredity. See, e.g., \citet{bien2013lasso,zhao2009composite} and \citet{haris2016convex}.

The standard errors for the estimated blip parameters can be obtained directly; a sandwich formula for computing the covariance of the estimates of the non-zero components can be derived \citep{SCAD}. How to derive the standard errors for the estimated blip parameters under the use of refitted pdWOLS requires further investigation. Post selection inference \citep{lee2016exact} should also be addressed. 


The proposed method is, fundamentally, based on prediction, selecting any variables that can improve predictive ability. As such, in finite samples, pdWOLS may underestimate the importance of variables that have small predictive ability but that play a significant role in DTRs.
Besides, the application of predictive methods directly to causal models may result in inflated variances and self-inflicted bias \citep{causal}. The importance of the distinction between DTRs (causal inference) and prediction must be kept in mind. Variable selection in causal inference is a tough problem: on the one hand, we want to adjust for enough covariates in the analysis to achieve ignorability; on the other hand, adjustment for some other irrelevant variables could induce bias and losses of statistical efficiency \citep{overad}. Hence, a thoughtful selection of confounders is needed, using expert knowledge to guide variable selection is encouraged. Other discussions about confounder selection can be found in \cite{susan,robinsgreenland,schneeweiss}. For pdWOLS, if we are worried about confounding and our focus is on building simple rules, we may want to do minimal selection on main effects but lots of selection on interaction effects, which can be implemented by setting small adaptive weights $w_j$ for the main effects or setting  $\alpha$ to a large value. How to choose the tuning parameter $\lambda$ and $\alpha$ in a DTR framework is an open and intriguing problem worthy of further investigation. 

\section*{Acknowledgements}
Research reported in this publication was supported by the National Institute of Mental Health of the National Institutes of Health under Award Number R01 MH114873 (co-PIs Shortreed and Moodie). The content is solely the responsibility of the authors and does not necessarily represent the official views of the National Institutes of Health. Moodie is a Canada Research Chair (Tier 1) in Statistical Methods for Precision Medicine and acknowledges the support of a chercheur de mérite  career award from the Fonds de Recherche du Québec, Santé. Bhatnagar acknowledges funding via a Discovery Grant from the Natural Sciences and Engineering Research Council of Canada (NSERC), RGPIN-2020-05133. Dr. Shortreed has been a co-Investigator on Kaiser Permanente Washington Health Research Institute projects funded by Syneos Health, who was representing a consortium of pharmaceutical companies carrying out FDA-mandated studies regarding the safety of extended-release opioids.

\bibliographystyle{apalike} 
\bibliography{bib.bib}

\section*{Supporting Information}

A Web Appendix containing the algorithm referenced in Section 2.4, regularity conditions and proofs of Theorems in Section 2.6, additional simulation results in Sections 3.2, 3.4, STAR*D details in Section 4 and an example of pdWOLS implemented in the R programming language are available with this paper at the Biometrics website on Wiley Online Library.\vspace*{-8pt}

\end{document}